\documentstyle[twoside,fleqn,espcrc2]{article}
\def\gsim{ \vcenter{\hbox{$\buildrel{\displaystyle >}\over\sim$}} }
\def\ema {e^{am}}
\def\Zax {{\zeta_{A}}}
\def\Zaxstat {{\hat{\zeta}_{A}}}
\def\Zxx {{\zeta_{B}}}
\def\Zxxstat {{\hat{\zeta}_{B}}}
\def\Gax {{G_{A}}}

\def\Gxx {{G_{B}}}

\def\K {{\kappa}}

\def\sigdotB {{\vec\sigma\cdot\vec B}}
\def\tm {{\tilde m}}

\def\chisqdof {{\chi^2/{\rm dof}}}

\def\bspf {{$\beta\!=\!6.4$}}
\def\bspthr {{$\beta\!=\!6.3$}}
\def\bsptwo {{$\beta\!=\!6.2$}}

\def\bsix {{$\beta\!=\!6.0$}}

\def\ainv {{a^{-1}}}

\newcommand{\AmS}{{\protect\the\textfont2
  A\kern-.1667em\lower.5ex\hbox{M}\kern-.125emS}}

\title{Quenched Lattice Results for $f_B$ and $f_D$}

\author{Claude Bernard,\address{Department of Physics,
	Washington University, St. Louis, MO 63130, USA}
  James Labrenz,\address{Department of Physics FM-15,
	University of Washington, Seattle, WA 98195, USA}%
	\thanks{Talk at Lattice '92, Amsterdam, 9/92, to be published in the
	proceedings. Preprint UW/PT-92-21, Wash. U. hep-lat/9211048
	(November 1992).}
        and
  Amarjit Soni\address{Department of Physics,
	Brookhaven National Laboratory, Upton, NY 11973, USA}}

\begin{document}

\begin{abstract}
We have computed the decay constants for the $B$ and $D$ mesons,
using quenched lattices at \bspthr, by interpolating between the
static approximation of Eichten and
the conventional (``heavy'' Wilson fermion) method.
A more careful treatment of the static result using
better sources with longer time-displacements and
a modification to the Wilson quark normalization to
correct approximately for lattice effects in the large-$am$ regime have led
to the elimination of the large discrepancy between the two methods
which had been previously observed.
We report final results, with estimates of various systematic errors.
\end{abstract}

\maketitle

\section{INTRODUCTION}

Last year we reported preliminary results \cite{lat91} for the heavy-light
decay constants $f_B$ and $f_D$ from wall-source lattices at \bspthr.
We found that systematic effects contributed to
the previously-observed discrepancy \cite{fB-wupp,fB-elc,lat90}
between the large-mass extrapolation of ``conventional''
lattice calculations (i.e., two propagating Wilson quarks)
and the static method of Eichten \cite{eichten-lat88}.

In this paper, we present our final results for this calculation
and focus primarily on the issues which led to the elimination
of the discrepancy.
First, we discuss briefly the way in which Wilson quarks have
been adapted for use when the mass is near or larger
than the lattice cut-off (i.e., where $am\gsim1$).
Second, we present a more comprehensive study of the static
computation at \bspthr\ and attempt to reduce
higher-state contamination in this result.
To compute the physical $B$ and $D$ mesons,
we interpolate between these two methods; our results
include estimates of various possible systematic errors.

\section{METHODS}

The calculation of the heavy-light decay constant is best
formulated in terms of the large-mass scaling quantity
$f_P\sqrt{M_P}$, where we use $P$ to refer to
a pseudoscalar of arbitrary heavy- and light-quark masses.
We compute it using the formula
\begin{equation}\label{fplatt}
	\phi_P \equiv f_P\sqrt{M_P} =

C_A\phantom{;}\left[2\zeta^{2}_{A}/\Zxx\right]^{1/2}\phantom{;}a^{-3/2}.
\end{equation}
The quantities $\Zax$ and $\Zxx$ are the fitted
large-time amplitudes of the two correlation functions, $\Gax$,
the ``smeared-point,'' and $\Gxx$, the ``smeared-smeared.''
(Here ``point'' refers to the local lattice axial current
and ``smeared'' to an extended pseudoscalar interpolating operator
defined in Coulomb gauge.)
$C_A$ normalizes the lattice current to the continuum one,
i.e., $A^{\rm cont}_\mu = C_A A^{\rm latt}_\mu$.
For the conventional method, we use
\begin{equation}\label{CA}
   C_A(\kappa_b,\kappa_q) =
        Z_A\sqrt{2\kappa_be^{a\tilde m_b}} \sqrt{2\kappa_qe^{a\tilde m_q}},
\end{equation}
where, for example,
\begin{equation}\label{am}
	a\tilde{m}_b = \ln\left(1 + {1\over u_0}({1\over 2\kappa_b}
		- {1\over 2\kappa_{c}})\right);\ u_0\equiv1/8\K_c.
\end{equation}
Here $b,q$ refers to the heavy, light quark.

Comparison to the spatially summed free-quark
propagator gives the quark-field normalization at leading
order in large $am$ \cite{large-am,NRQCD}. One expects corrections
of order $\alpha_s am$ and of order $1/M_P$;
the former can be further reduced by
tadpole improvement \cite{large-am,lepmac-lat90,NRQCD},
embodied in the use of $u_0\K$ and $u_0\K_c$,
instead of $\K$ and $\K_c$, in (\ref{am}).
We attempt to estimate the size of the latter below.
Note that only in the factor $e^{a\tilde m}$
is the tadpole approximation used.
For the leading perturbative correction
to the axial current we retain the full result \cite{ZA},
$Z_A = 1 - .133g^2$, and use a ``boosted'' value of the coupling
($g^2=1.62$ at \bspthr) \cite{lepmac-lat90}.

For the static calculation, we extract residues of the bare
correlation functions, $\Zaxstat$ and $\Zxxstat$, to determine
$\hat\phi_P$ in a formula analogous to (\ref{fplatt}).
Here we use the normalization constant \cite{boucaud,hill2}
$$\displaylines{
\hat C_A = \hat{Z}_A C(a,m_b)\sqrt{2\K_qe^{a\tilde m_q}},\hfill\cr
\hat{Z}_A = 1 - 22.38\lambda,\ \ \lambda=g^2C_F/16\pi^2,\hfill\cr
    C(a,m) = 1 + 3\lambda\ln(am).\hfill}
$$
This form is convenient for later purposes.

Note that Equation (\ref{CA})
produces the usual normalization for the decay constant when
{\it both} quarks are light, since the factors of $\ema\rightarrow1$.
Furthermore, it matches to the static limit (up to perturbative corrections),
since, in the tadpole approximation,
$Z_A\sqrt{2\K_be^{a\tilde m_b}} \approx 1$ as $\K_b\rightarrow 0$.

Consider, though, the error made in the intermediate mass regime,
where one hopes to obtain the $1/M$ correction to the static limit.
In a hopping expansion for the
heavy quark, one finds that $M_P\rightarrow \ln(1/2\K_b)$,
and that sub-leading terms enter with additional powers of $\K$.
Thus from the standpoint of NRQCD \cite{NRQCD}, the $1/M$ corrections to the
static term will be exponentially suppressed, and the relative
importance of the kinetic and $\sigdotB$ contributions will be misrepresented.

We improve the situation as follows.
In the non-relativistic limit,
at arbitrary $am$, the
(tadpole improved) free-quark dispersion relation,
$E(\vec{k}) = \tilde m + \vec{k}^2/2\tm_2 + \dots$,
contains the kinetic mass
$$
        a\tm_2 = (e^{a\tm}\sinh{a\tm})/(\sinh{a\tm} +1).
$$
In this limit,
the rest mass $\tm$ is irrelevant in the calculation of $\phi_P$,
but it clearly contributes to the pole mass, $M_P$.
Thus, assuming that non-tadpole-interaction effects are sub-dominant,
the shifted value
\begin{equation}\label{Mshift}
        M_P \rightarrow M_P + (\tilde m_2 - \tilde m)
\end{equation}
will correspond to an amplitude $\phi_P$ where the
heavy-quark kinetic term is properly included.

Unfortunately, we cannot simultaneously correct
for $\sigdotB$ interactions without a more
sophisticated approach---i.e., a modification to the action
and the tuning of an additional parameter \cite{kronfeld}.
As the best approximation that we can make within these limitations,
we use the shift described by (\ref{Mshift}) in our analysis but
include its full effect
on the final results as an additional systematic error.
Its effect is actually quite small.

\section{NUMERICAL RESULTS}

For the static limit at \bspthr,
we have used both wall sources \cite{lat91} and ``cube''
sources (see, for example, \cite{fB-elc,lat90}).
The latter are constructed, also in Coulomb gauge, from
point-source propagators, and the smearing volume can be adjusted.

\begin{figure}[tb]
\vspace{2.2truein}
\includegraphics{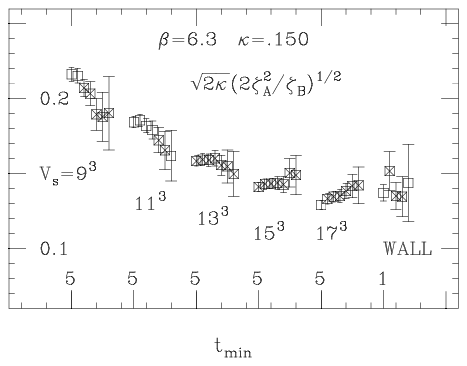}
\caption{Static dependence on $V_s$ and $t_{\rm min}$}
\end{figure}

In Fig. 1 we show the results of the static decay
amplitude for various smearing volumes.
For each volume, we show in addition the dependence
on the fitted time interval of the correlator $\Gax$, while that
for $\Gxx$ has been held fixed.
Each analysis consists of a coupled fit (i.e., both correlators
fit simultaneously, extracting a single ground-state mass)
and includes correlations in the data.
We mark the fits which satisfy $\chisqdof<1$ with ``$\times$.''

A search for an early plateau in the effective masses,
$M_A(t)$ and $M_B(t)$, or in $G_A/G_B$,
can lead to some judgement of the ``best'' smearing
volume to use. However, the following difficulties arise:
(1) the possibility of short-range false plateaus
in $M_A$ coupled with a decreasing signal-to-noise ratio at large
times;
(2) the location of the  plateaus in $M_A(t)$, $M_B(t)$,
and  $G_A/G_B$ may not coincide,
making the ratio $G_A/G_B$ difficult to interpret unambiguously;
and
(3) a noticeable variation in the optimum volume as the
light-quark mass is changed complicates extrapolations to $\K_c$.

Fig. 1 provides an example where reasonable
fits are obtained at early times for three different smearing volumes,
namely $V_s=13^3$, $15^3$, and $17^3$, but where systematic differences
are clearly evident in the raw amplitude.
However, the later-time fits for these three smearing
sizes are in good agreement, and they
show consistency with the wall source results which were extracted
using a different method, as discussed below.
With these rather crude smearing techniques, and with our level of statistics,
we view the safest choice for a quoted result to be a later-time
fit where the choice of the ``optimum'' volume is less of an issue,
since here we obtain consistent results over a range of smearing sizes,
including the wall sources.  Similar analysis
of the static computations at $\beta=6.0$ \cite{lat91} gave a considerably
smaller value for $f_P\sqrt{M_P}$  than initially reported.

Note that we do {\it not} expect
that the projection onto the ground state will continue to
improve as we enlarge the smearing volume.
As we have previously pointed
out \cite{lat91}, the relative higher-state contamination is expected
to increase due to the large extent of the wall source.
However, the wall has  an advantage in terms of statistics
(sources for both light and heavy quarks at every point on a timeslice),
and ultimately its performance must be evaluated empirically.
We have indeed found a two-state fit necessary for the wall sources,
indicating a substantial admixture of an excited state.
Nonetheless we find good agreement
(both in central value and size of errors) in the ground-state quantities
with the best cube-source results.
Since we view a single-state fit as a safer alternative,
we have now chosen the $V_s=15^3$ source (presented in table 1)
for our combined analysis leading to $f_B$ and $f_D$.
Due to the limitations of our sources,
we cannot rule out a systematic error in the static results
of roughly $10\%$.

\begin{figure}[tb]
\vspace{2.2truein}
\includegraphics{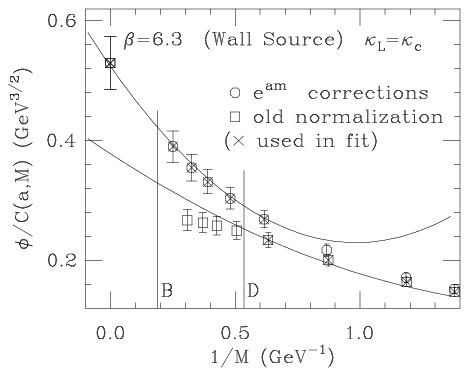}
\caption{Analysis of large-$am$ systematics}
\end{figure}

In Fig 2, we show the results of two analyses combining
the static point ($V_s=15^3$ source, single-state fit)
and the conventional data (wall-source, two-state fit),
both with and without the large-$am$ corrections.
The light quark has been extrapolated to the chiral limit.
The analysis procedure is:
(1) divide the conventional amplitudes by $C(a,m)$;
(2) fit to a quadratic in $1/M$ and interpolate to the $D$ and $B$ masses;
(3) re-introduce $C(a,m)$ in the interpolated results.
The fit is made using a jackknife estimation of the covariance matrix to
account for correlations.

When the $\ema$ normalization is used, we obtain good fits using
the heaviest of the conventional data ($\chisqdof=1.9/3$ for
the chiral extrapolation, $\chisqdof=3.3/3$ for the strange).
If we attempt to fit the same set of points with the ``old''
normalization, or even if we throw out the heaviest points,
thereby attempting to avoid large-$am$ errors, we obtain much poorer
fits. (The latter option is shown as ``old normalization'' in Fig.~2.
The fit has $\chisqdof=8.2$.)

These data strongly indicate that the $\ema$ normalization is necessary
to get agreement between the two methods.
Although we cannot rigorously bound the
systematic error associated with the approximate techniques which
we have used, we expect that a rough estimate will be given by
the size of the change due to (\ref{Mshift}),
since this directly addresses $1/M$ effects in a region where we
expect the heavy quark to behave non-relativistically.
However, since the corrections to the lighter masses are small,
and the static point remains unchanged, the difference in the fit
with and without (\ref{Mshift}) is minimal, as reflected
in the ``$1/M$'' values in tables 2 and 3.

\begin{table}[t]
\caption{Static results, $V_s=15^3$}
\label{tab-static}
\begin{tabular}{cccc}
\hline
$\K$ & $a{\cal E}$ & $\hat{C}_A/C(a,m)$ & $a^{3\over2}\hat{\phi}\times 10$ \\
\hline
.149 & 0.525(10) & 0.392 & 1.15(6) \\
.150 & 0.510(11) & 0.388 & 1.07(6) \\
.1507 & 0.499(11)& 0.385 & 1.00(6) \\
$\K_s$  & 0.503(11)& - &  1.02(6) \\
$\K_c$  & 0.486(12)& - &  0.92(6) \\
\hline
\end{tabular}
\end{table}

Uncertainties due to the choice of coupling used in the
perturbative corrections (labeled ``$g_0$'' in the tables)
are obtained by re-analyzing the data using
the bare value, $g_{0}^2 = 6/\beta$.
This error is also small because we compute $f_\pi$ to
determine the scale; thus the results are insensitive to $Z_A$.
With a boosted coupling we find $\ainv\!=\!3.2(1)$ GeV.

We have made a similar analysis at \bsix, but our statistics
for the heaviest masses and the static point are quite poor.
Since we are therefore unable to reliably compute a scaling error
by extrapolating to $a=0$, we crudely estimate remaining systematics
by re-analyzing the data with the scale shifted by $\pm15\%$.
We obtain this value by interpolating the comparison of
$(\ainv)_{f_\pi}$ and $(\ainv)_{m_\rho}$ made at
\bsptwo\ and \bspf\ in Ref.~\cite{elc-6.4}.
This provides a conservative estimate; using $f_K$ to set the
scale, for example,
we find a shift of only $8\%$.

\section{CONCLUSIONS}

The use of better sources and longer time displacements
for the static method, and the modification of the Wilson quark
normalization for large $am$, now give consistency between the
static and conventional methods.
Without such corrections, we find
that a significant discrepancy still exists, even at \bspthr.
Results are given in tables 1--3.

Further study of systematics is certainly warranted.
Improved sources for both methods would help to
eliminate potential systematic errors from excited state contamination.
A more rigorous treatment of the heavy quark is
necessary in order to better understand the large-$am$ errors which
we have attempted to estimate here. This would then facilitate
studies of other key issues, such as scale violations.

\begin{table}[t]
\caption{Decay constants and systematic error estimates}
\label{tab-results}
\begin{tabular}{ccccccc}
\hline
      & $f$ (MeV) &     & fits & $g_0$ & $1/M$ & $\ainv$ \\
\hline
$B$   &  187(10)  & $\pm$ & 12 & 3 & 4  & 37   \\
$B_s$ &  207(9)	 & $\pm$ & 10 & 3 & 5  & 39   \\
$D$   &  208(9)	 & $\pm$ & 11 & 1 & 7  & 34   \\
$D_s$ &  230(8)	 & $\pm$ & 10 & 9 & 6  & 34   \\
\hline
\end{tabular}
\end{table}

\begin{table}[t]
\caption{Jackknifed Ratios}
\label{tab-ratios}
\begin{tabular}{lllllll}
\hline
      &  &     & fits & $g_0$ & $1/M$ & $\ainv$ \\
\hline
$f_B/f_D$         & $\!\!.90(3)$ & $\!\pm\!$ & .02 & .02 & $<.01$    & .04   \\
$f_{B_s}/f_{D_s}$ & $\!\!.90(2)$ & $\!\pm\!$ & .02 & .02 & $<.01$    & .05   \\
$f_{B}/f_{B_s}$   & $\!\!.90(2)$ & $\!\pm\!$ & .03 & .03 & $\sim.01$ & .02   \\
$f_D/f_{D_s}$     & $\!\!.90(2)$ & $\!\pm\!$ & .02 & .03 & $<.01$    & .02   \\
\hline
\end{tabular}
\end{table}

\section*{ACKNOWLEDGEMENTS}
We thank E.~Eichten, B.~Hill, A.~Kronfeld,
P.~Lepage, P.~Mackenzie, and S.~Sharpe
for many useful conversations.
J.L. would like to thank the theory group at
Brookhaven National Laboratory for support while this
work was in progress.
Computing was done at the San Diego Supercomputer Center and at
the National Energy Research Supercomputer Center.
This work was supported in part by the DOE under grant numbers
DE-2FG02-91ER40628 (C.B.) and DE-AC02-76CH00016 (J.L. and A.S.).

%

\def\PRL#1#2#3{{Phys. Rev. Lett.} {\bf #1}, #3 (#2) }
\def\PRD#1#2#3{{Phys. Rev.} {\bf D#1}, #3 (#2)}
\def\PLB#1#2#3{{Phys. Lett.} {\bf #1B} (#2) #3}
\def\NPB#1#2#3{{Nucl. Phys.} {\bf B#1} (#2) #3}
\def\NPBPS#1#2#3{{Nucl. Phys.} {\bf B ({Proc. Suppl.}){#1}} (#2) #3}
\def\etal {{\it et al.}}


\end{document}